\PassOptionsToPackage{hyphens}{url}

\documentclass[10pt,twocolumn]{article}

\pdfoutput=1
\usepackage{times}
\usepackage{fullpage}
\usepackage{graphicx}
\usepackage{epstopdf}

\begin{document}

\title{\bf Lightweight User-Space Record And Replay}
\author{Robert O'Callahan, Chris Jones, Nathan Froyd, Kyle Huey, Albert Noll, Nimrod Partush}
\date{}
\maketitle
\thispagestyle{empty}

\newcommand{\system}{\textsc{rr}}

\begin{abstract}
The ability to record and replay program executions with low overhead
enables many applications, such as reverse-execution debugging,
debugging of hard-to-reproduce test failures, and ``black box''
forensic analysis of failures in deployed systems. Existing
record-and-replay approaches rely on recording an entire virtual machine
(which is heavyweight), modifying the OS kernel (which adds deployment and maintenance costs),
or pervasive code instrumentation (which
imposes significant performance and complexity overhead). We investigated
whether it is possible to build a practical record-and-replay system avoiding all
these issues. The answer turns out to be yes --- if the CPU and
operating system meet certain
non-obvious constraints. Fortunately modern Intel CPUs, Linux
kernels and user-space frameworks meet these constraints, although this has only become true recently. With
some novel optimizations, our system \system{} records and replays real-world workloads
with low overhead with an entirely user-space implementation running on stock hardware
and operating systems. \system{} forms the basis
of an open-source reverse-execution debugger seeing significant use in practice.
We present the design and implementation of \system{}, describe its performance on a
variety of workloads, and identify constraints on hardware and operating system design required
to support our approach.
\end{abstract}

\section{Introduction}

The ability to record a program execution with low overhead and play it back
precisely has many applications \cite{Devecsery2014,Dolan-Gavitt2015,Engblom2012} and has received significant attention
in the research community. It has even been implemented in products such as VMware Workstation \cite{Malyugin2007}, Simics \cite{Engblom2010}, UndoDB \cite{UndoDB} and TotalView \cite{Gottbrath2008}.
Unfortunately, deployment of these techniques has been limited, for
various reasons. Some approaches \cite{Dunlap2002,Engblom2010,Malyugin2007} require recording and replaying an entire
virtual machine, which is heavyweight. Other approaches \cite{Bergan2010,Devecsery2014,Laadan2010} require running
a modified OS kernel, which is a major barrier for users
and adds security and stability risk to the system. Some approaches \cite{Hower2008,Montesinos2008,Pokam2013} require custom hardware not yet available. Many approaches \cite{UndoDB,Bhansali2006,Gottbrath2008,Patil2010} require
pervasive instrumentation of code, which adds complexity and overhead, especially for self-modifying code (commonly used in polymorphic inline caching \cite{Holzle1991} and other implementation techniques in modern just-in-time compilers). A performant dynamic code instrumentation engine is also expensive to build and maintain.

We set out to build a system that maximizes deployability by avoiding all these issues: to record and replay
unmodified user-space applications on stock Linux kernels and x86/x86-64 CPUs, with a fully user-space implementation running without special
privileges, and without using pervasive code instrumentation. We assume that \system{} should run unmodified applications, and they will have bugs (including data races) that we wish to faithfully record
and replay, but these applications will not maliciously try to subvert recording or replay.
We combine techniques already known, but not previously demonstrated working together in a practical
system: primarily, using {\tt ptrace} to record and replay system call results and signals,
avoiding non-deterministic data races by running only one thread at a time,
and using CPU hardware performance counters to measure application progress so asynchronous signal
and context-switch events are delivered at the right moment \cite{Olszewski2009}.
Section \ref{design} describes our approach in more detail.

With the basic functionality in place, we discovered that the main performance bottleneck for
our applications was the
context switching required by using {\tt ptrace} to monitor system calls. We implemented
a novel {\it system-call buffering} optimization to eliminate those context switches, dramatically
reducing recording and replay overhead on important real-world workloads. This optimization relies on modern Linux kernel
features: {\tt seccomp-bpf} to selectively suppress {\tt ptrace} traps for certain system
calls, and {\tt perf} context-switch events to detect recorded threads blocking in the kernel.
Section \ref{syscallbuf} describes this work, and Section \ref{results} gives some performance
results, showing that on important application workloads \system{} recording and replay slowdown is less than a factor of two.

We rely on hardware and OS features designed for other goals, so it is surprising that
\system{} works. In fact, it skirts the edge of feasibility, and in particular
it cannot be implemented on ARM CPUs. Section \ref{constraints} summarizes \system{}'s hardware and software requirements, which we hope will influence system designers.

\system{} is in daily use by many developers as the foundation of an efficient reverse-execution debugger that works on complex applications such as Samba, Firefox, QEMU, LibreOffice and Wine. It is free software, available at \url{https://github.com/mozilla/rr}. This paper makes the following research contributions:
\begin{itemize}
\item We show that record and replay of Linux user-space processes
on modern, stock hardware and kernels and without pervasive code instrumentation is possible and practical.
\item We introduce the {\it system-call buffering} optimization and show that it dramatically reduces overhead.
\item We show that recording and replay overhead is low in practice, for applications that don't use much parallelism.
\item We identify hardware and operating system design constraints required to support our approach.
\end{itemize}

\section{Design} \label{design}

\subsection{Summary}

Most low-overhead record-and-replay systems depend on the observation that CPUs are mostly
deterministic. We identify a boundary around state and computation, record all sources
of nondeterminism within the boundary and all inputs crossing into the boundary, and reexecute
the computation within the boundary by replaying the nondeterminism and inputs. If all inputs and
nondeterminism have truly been captured, the state and computation within the boundary during replay
will match that during recording.

To enable record and replay of arbitrary Linux applications, without requiring kernel modifications or
a virtual machine, \system{} records and replays the user-space execution of a group of processes. To simplify
invariants, and to make replay as faithful as possible, replay preserves almost
every detail of user-space execution. In particular, user-space memory and register values are preserved
exactly, with a few exceptions noted later in the paper. This implies CPU-level control flow is
identical between recording and replay, as is memory layout.

While replay preserves user-space state and execution, only a minimal amount of kernel state is
reproduced during replay. For example, file descriptors are not opened, signal handlers are not
installed, and filesystem operations are not performed. Instead the recorded user-space-visible effects of those
operations, and future related operations, are replayed. We do create one replay thread per
recorded thread (not strictly necessary), and we create one replay address
space (i.e.\ process) per recorded address space, along with matching memory mappings.

With this design, our recording boundary is the interface between user-space and the kernel.
The inputs and sources of nondeterminism are mainly the results of system calls, and the timing of
asynchronous events.

\subsection{Avoiding Data Races}

If we allow threads to run on multiple cores simultaneously and share memory, racing read-write or write-write accesses to
the same memory location by different threads would be a source of
nondeterminism. It seems impractical to record such nondeterminism with
low overhead on stock hardware, so instead we avoid such races by allowing only one recorded
thread to run at a time. \system{} preemptively schedules these threads, so context switches
are a source of nondeterminism that must be recorded. Data race bugs can still be observed if a context
switch occurs at the right point in the execution (though bugs due to weak memory models cannot be
observed).

This approach is simple and efficient for applications without much parallelism. It imposes
significant slowdown for applications with a consistently high degree of parallelism, but those are still relatively uncommon.

\subsection{System Calls} \label{syscalls}

System calls return data to user-space by modifying registers and memory, and these changes must be recorded. The {\tt ptrace}
system call allows a process to supervise the execution of other ``tracee'' processes and threads,
and to be synchronously notified when a tracee thread enters or exits a system call. 
When a tracee thread enters the kernel for a system call, it is suspended and
\system{} is notified. When \system{} chooses to run that thread again, the system call will complete, notifying
\system{} again, giving it a chance to record the system call results. \system{} contains a model of most Linux system calls
describing the user-space memory they can modify, given the system call input parameters and result.

As noted above, \system{} normally avoids races by scheduling only one thread at a time. However,
if a system call blocks in the kernel, \system{} must try to schedule another application thread to run
while the blocking system call completes. It's possible (albeit unlikely) that the running thread
could access the system call's output buffer and race with the kernel's writes to that buffer. To avoid this, we redirect
system call output buffers to per-thread temporary ``scratch memory'' which is otherwise unused by the application.
When we get a {\tt ptrace} event for a blocked system call completing, \system{} copies scratch buffer
contents to the real user-space destination(s) while no other threads are running, eliminating the race.

\begin{figure}[t]
\centering
\includegraphics[scale=0.4]{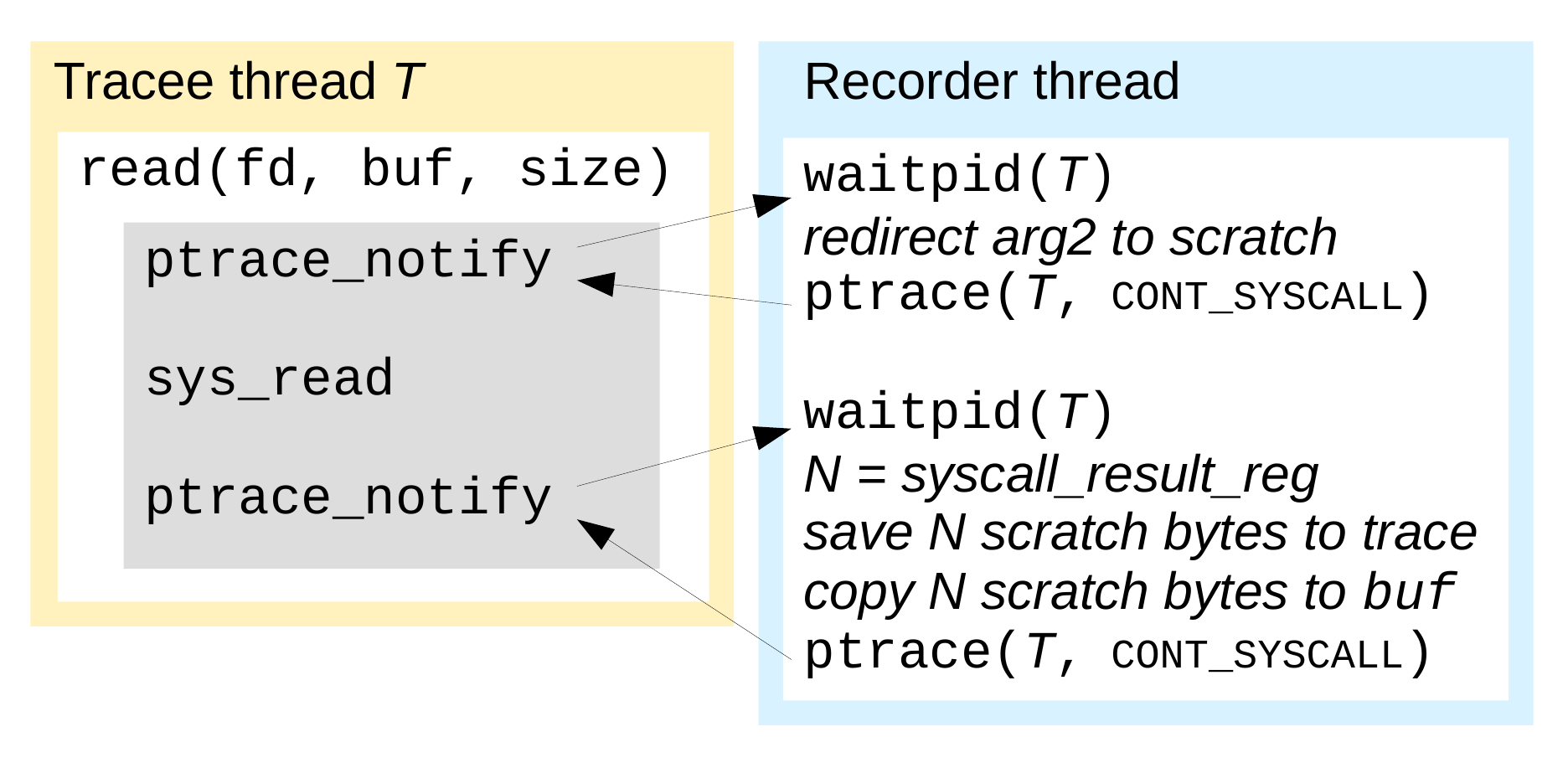}
\caption{Recording simple system call}
\label{simple-syscall}
\end{figure}

Figure \ref{simple-syscall} illustrates the flow of control recording a simple {\tt read} system call. The gray box represents kernel code.

During replay, when the next event to be replayed is an intercepted system call, we
set a temporary breakpoint at the address of the system call instruction (recorded in the trace). We use
{\tt ptrace} to run the tracee thread until it hits
the breakpoint, remove the breakpoint, advance the program counter past the system call instruction,
and apply the recorded register and memory changes. This approach is simple and minimizes the number
of context switches between \system{} and the tracee thread. (Occasionally it is unsafe and we fall back
to a more complicated mechanism.)

\subsection{Asynchronous Events}

We need to support two kinds of asynchronous events: preemptive context switches and signals.
The former is handled as a special case of the latter; we force a context switch by sending a signal
to a running tracee thread. In general we need to ensure that during replay, a signal is delivered when the
program is in exactly the same state as it was when the signal was delivered during recording.

Our approach is very similar to ReVirt's \cite{Dunlap2002}: use CPU hardware performance counters. Ideally we would count the number of
retired instructions leading up to a signal delivery during recording, and during replay, program
the CPU to fire an interrupt after that many instructions have been retired. However, this simple approach needs
two modifications to work in practice.

\subsubsection{Nondeterministic Performance Counters} \label{counters}

Measuring user-space progress with a hardware performance counter for record-and-replay requires that every time we execute
a given sequence of user-space instructions, the counter value changes by an amount that depends only
on the instruction sequence, and not any system state unobservable from user space (e.g.\ the contents of
caches, the state of page tables, or speculative CPU state). As noted in previous work \cite{Dunlap2002,Weaver2013}, this property (commonly described as ``determinism'') does not hold for
most CPU performance counters in practice. For example it does not hold for any ``instructions retired''
counter on any known x86 CPU model, because (among other reasons) an instruction triggering a page fault is restarted and counted twice.

Fortunately, modern Intel CPUs have exactly one deterministic performance counter: ``retired conditional branches'' (``RCB''), so
we use that. We cannot just count the number of RCBs during recording and deliver the signal after we have
executed that number of RCBs during replay, because the RCB count does not uniquely determine the execution point to
deliver the signal at. Therefore we pair the RCB count with the complete state of general-purpose registers (including the program counter)
to identify an execution point. In theory that \emph{still} does not uniquely identify an
execution point: consider, for example, the infinite loop {\tt label: inc [global\_var]; jmp label;}. In practice applications do not do that sort of thing and we have found it works reliably.

\subsubsection{Late Interrupt Firing}

The other major problem is that, although CPUs can be programmed to fire an interrupt after a specified number of
performance events have been observed, the interrupt does not fire immediately. In practice we often observe it firing after
dozens more instructions have retired. To compensate for this, during replay, we program the interrupt to trigger
some number of events earlier than the actual RCB count we are expecting. Then we set a temporary breakpoint
at the program counter value for the state we're trying to reach, and repeatedly run to the breakpoint until the RCB count and the general-purpose register values match their recorded values.

\subsection{Shared Memory} \label{shmem}

By scheduling only one thread at a time, \system{} avoids issues with races on shared memory as long as
that memory is written only by tracee threads. It is possible for recorded processes to share memory
with other processes, and even kernel device drivers, where that non-recorded code can perform writes
that race with accesses by tracee threads. Fortunately, this is rare for applications running in common Linux desktop environments, occurring in only four common cases: applications sharing memory with the PulseAudio daemon,
applications sharing memory with the X server, applications sharing memory with kernel graphics drivers and GPUs, and {\tt vdso} syscalls. We avoid the first three problems by
automatically disabling use of shared memory with PulseAudio and X (falling back to a socket transport in both cases), and disabling direct access to the GPU from applications.

{\tt vdso} syscalls are a Linux optimization that implements some common read-only system calls (e.g.\ {\tt gettimeofday}) entirely in user space, partly by reading memory shared with the kernel and
updated asynchronously by the kernel. We disable {\tt vdso} syscalls by patching their user-space implementations
to perform the equivalent real system call instead.

Applications could still share memory with non-recorded processes in problematic ways, though this is rare in practice and
can often be solved just by enlarging the scope of the group of processes recorded by \system{}.

\subsection{Nondeterministic Instructions} \label{instructions}

Almost all CPU instructions are deterministic, but some are not. One common nondeterministic x86 instruction
is {\tt RDTSC}, which reads a time-stamp counter. This particular instruction is easy to handle, since the
CPU can be configured to trap on an {\tt RDTSC} and Linux exposes this via a {\tt prctl} API, so we can
trap, emulate and record each {\tt RDTSC}.

Other relatively recent x86 instructions are harder to handle. {\tt RDRAND} generates random numbers and hopefully is
not deterministic. We have only encountered it being used in one place in GNU {\tt libstdc++}, so \system{} patches that explicitly. {\tt XBEGIN} and associated instructions
support hardware transactional memory. These are nondeterministic from the point of view of user space, since
a hardware transaction can succeed or fail depending on CPU cache state. Fortunately so far we have only
found these being used by the system {\tt pthreads} library, and we dynamically apply custom patches to that
library to disable use of hardware transactions.

The {\tt CPUID} instruction is mostly deterministic, but one of its features returns the index of the
running core, which affects behavior deep in {\tt glibc} and can change as the kernel
migrates a process between cores. We use the Linux {\tt sched\_setaffinity} API to force all tracee threads to run
on a particular fixed core, and also force them to run on that core during replay.

We could easily avoid most of these issues in well-behaved programs if we could just trap-and-emulate the {\tt CPUID}
instruction, since then we could mask off the feature bits indicating support for {\tt RDRAND}, hardware
transactions, etc. Modern Intel CPUs support this (``{\tt CPUID} faulting''); we are in the process of adding an API for this to Linux.

\subsection{Reducing Trace Sizes}

For many applications the bulk of their input is memory-mapped files, mainly executable code. Copying
all executables and libraries to the recorded trace on every execution would impose significant time and space overhead.
\system{} creates hard links to memory-mapped executable files instead of copying them; as long as a system update
or recompile replaces executables with new files, instead of writing to the existing files, the links retain the
old file data. This works well in practice.

Even better, modern filesystems such as XFS and Btrfs offer copy-on-write logical copies of files (and even block ranges within files), ideal for our purposes. When a mapped file is on the same filesystem as the recorded trace, and the filesystem supports cloning, \system{} clones mapped files into the trace. These clone operations are essentially free in time and space, until/unless the original file is modified or deleted.

\system{} compresses all trace data, other than cloned files and blocks, with the {\tt zlib} ``deflate'' method.

With these optimizations, in practice trace storage is a non-issue. Section \ref{space-results} presents some results.

\subsection{Other Details}

Apart from those major issues, many other details are required to build a complete record-and-replay system, too many
to mention here. Some system calls (e.g.\ {\tt execve}) are especially complex to handle. Recording and
replaying signal delivery are complex, partly because signal delivery has poorly-documented side effects on
user-space memory. Advanced Linux kernel features such as {\tt unshare}
(kernel namespaces) and {\tt seccomp} require thoughtful handling. Many of these details are interesting, but they do not impact
the overall approach.

\section{System Call Buffering} \label{syscallbuf}

The approach described in the previous section works, but overhead is disappointingly high (see Figure \ref{optimizations-chart} below). The core problem is that for every tracee system call,
as shown in Figure \ref{simple-syscall} the tracee performs four context switches: two blocking {\tt ptrace} notifications, each requiring a context switch from the tracee to \system{} and back. For common system calls such as {\tt gettimeofday} or {\tt read} from cached files,
the cost of even a single context switch dwarfs the cost of the system call itself. To significantly reduce overhead, we must
avoid context-switches to \system{} when processing these common system calls.

Therefore, we inject into the recorded process a library that intercepts common system calls, performs the system call without triggering
a {\tt ptrace} trap, and records the results to a dedicated buffer shared with \system{}. \system{} periodically flushes the buffer to its
trace. The concept is simple but there are problems to overcome.

\subsection{Intercepting System Calls}

When the tracee makes a kernel-entering system call, \system{} is notified via a {\tt ptrace} trap. It tries to
rewrite the system-call instruction to call into the interception library instead. This is tricky
because on x86 a system call instruction is two bytes long, but we need to replace it with a five-byte {\tt call}
instruction. (On x86-64, to ensure we can call from anywhere in the address space to the interception
library, we also need to also allocate trampolines within 2GB of the patched code.) The vast majority of
executed system call instructions are followed by one of only five different instructions;
for example, very many system call instructions are followed by a particular {\tt cmp} instruction testing
the syscall result. So we added five hand-written stubs to our interception library that
execute those post-system-call instructions before returning to the patched code, and on receipt of a {\tt ptrace} system-call notification, \system{}
replaces the system call instruction and its following instruction with a call to the corresponding stub.

For simplicity we (try to) redirect all system call instructions to the interception library,
but the library only contains wrappers for the most frequently called system calls. For other system
calls it falls back to doing a regular {\tt ptrace}-trapping system call.

\subsection{Selectively Trapping System Calls}

The classic {\tt ptrace} system-call interception API triggers traps for all system calls, but our
interception library needs to avoid traps for selected system calls. Fortunately,
modern Linux kernels provide a facility for selectively generating {\tt ptrace} traps: {\tt seccomp-bpf}.
{\tt seccomp-bpf} was designed primarily for sandboxing. A process can apply a {\tt seccomp-bpf} filter function, expressed in bytecode, to another process; then, for every system call performed by the target process, the kernel runs the filter. The filter has
access to incoming user-space register values, including the program counter. The filter result directs the kernel to either allow the system call, fail with a given {\tt errno}, kill the target process, or trigger a {\tt ptrace} trap. The overhead of filter execution is
negligible since filters run directly in the kernel and are even compiled to native code on most architectures.

\begin{figure}[t]
\centering
\includegraphics[scale=0.4]{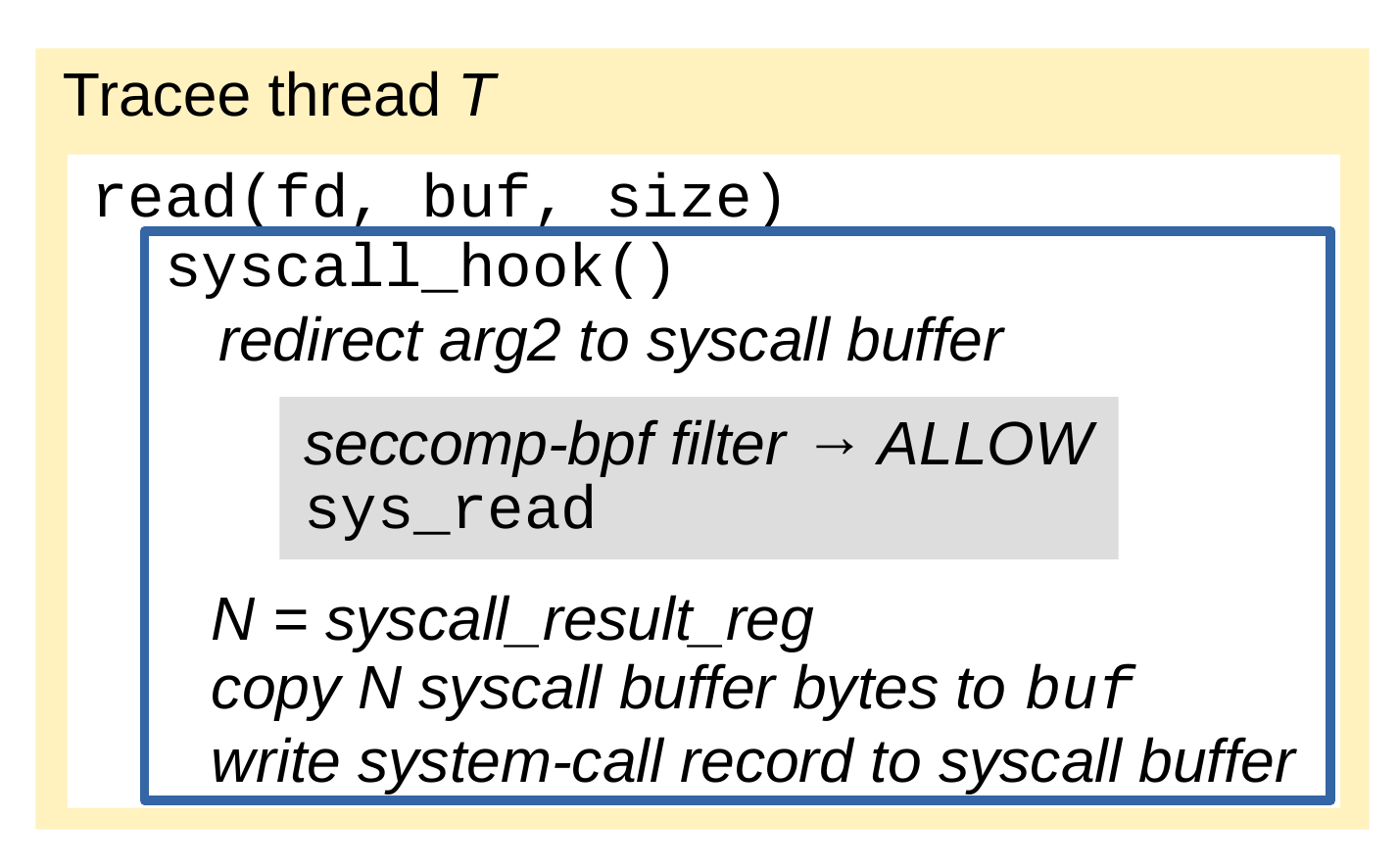}
\caption{Recording with system-call buffering}
\label{syscallbuf-diagram}
\end{figure}

Figure \ref{syscallbuf-diagram} illustrates recording a simple {\tt read} system call with system-call buffering. The solid-border box represents code in the interception library and the grey box represents kernel code.

\system{} injects a special page of memory into every tracee process at a fixed address (immediately after {\tt execve}).
That page contains a system call instruction --- the ``untraced instruction''.
\system{} applies a {\tt seccomp-bpf} filter to each recorded process that triggers a {\tt ptrace} trap for
every system call --- except when the program counter is at the untraced instruction, in which case
the call is allowed. Whenever the interception library needs to make an untraced system call, it uses that
instruction.

\subsection{Detecting Blocked System Calls}

Some common system calls sometimes block (e.g.\ {\tt read} on an empty pipe). Because \system{} runs
tracee threads one at a time, if a thread enters a blocking system call without notifying \system{}, it
will hang and could cause the entire recording to deadlock (e.g.\ if another tracee thread is about to
{\tt write} to the pipe). We need the kernel to notify \system{} and suspend the tracee thread whenever
it blocks in an untraced system call, to ensure we get a chance to schedule a different tracee thread.

We can do this using the Linux {\tt perf} event system to monitor {\tt PERF\_COUNT\_SW\_CONTEXT\_SWITCHES}.
One of these events occurs every time the kernel deschedules a thread from a CPU core. 
The interception library opens a file descriptor for each thread to monitor these events for that thread, and requests that the
kernel send a signal to the thread every time the event occurs. These signals trigger {\tt ptrace}
notifications to \system{} while preventing the thread from executing further.
To avoid spurious signals (e.g.\ when the thread is descheduled due to normal timeslice expiration), event
counting is normally disabled and explicitly enabled during an untraced system call that might block. It is still possible for spurious {\tt SWITCHES} to occur at any point between enabling and disabling the event; we handle these edge cases with careful inspection of the tracee state.

\begin{figure}[t]
\centering
\includegraphics[scale=0.4]{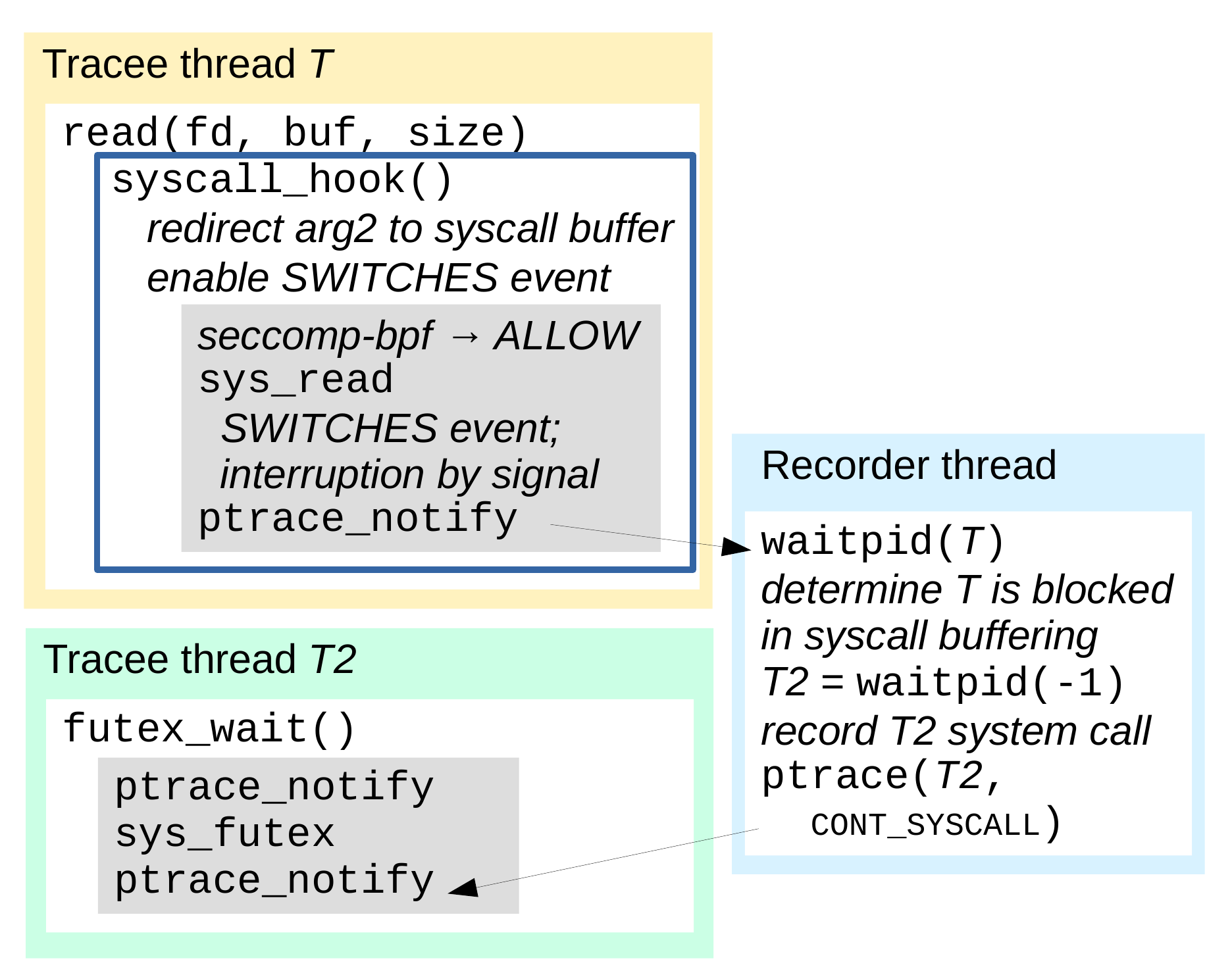}
\caption{Recording a blocking system call}
\label{syscallbuf-blocked-diagram}
\end{figure}

Figure \ref{syscallbuf-blocked-diagram} illustrates recording a blocking {\tt read} system call with system-call buffering. The kernel deschedules the thread, triggering a {\tt perf} event which sends a signal to the thread, \emph{rescheduling} it, interrupting the system call, and sending a {\tt ptrace} notification to the recorder. The recorder does bookkeeping to note that a buffered system call was interrupted in thread {\it T}, then checks whether any tracee threads in blocking system calls have progressed to a system-call exit and generated a {\tt ptrace} notification. In this example {\it T2} has completed a (not system-call-buffered) blocking {\tt futex} system call, so we resume executing {\it T2}.

Resuming a buffered system call that was interrupted by a signal (e.g.\ {\it T}'s {\tt read} call in Figure \ref{syscallbuf-blocked-diagram}) is more complicated and requires understanding Linux system call restart semantics, which are too nasty to cover in this paper.

\subsection{Handling Replay}

Conceptually, during recording we need to copy system call output buffers to a trace buffer, and during replay
we need to copy results from the trace buffer to system call output buffers. This is a problem
because the interception library is part of the recording and replay and therefore should
execute the same code in both cases.

For this reason (and because we want to avoid races of the sort discussed in Section \ref{syscalls}),
the interception library redirects system call outputs so that results are written directly to the trace buffer.
After the system call completes, the interception library copies the output data from the trace buffer to
the original output buffer(s). During replay the untraced system call instruction is replaced with a no-op, so the system call
does not occur; the results are already present in the trace buffer so the post-system-call copy from the
trace buffer to the output buffer(s) does what we need.

During recording, each untraced system call sets a result register and the interception library writes it to the
trace buffer. During replay we must read the result register from the trace buffer instead. We
accomplish this with a conditional move instruction so that control flow is perfectly consistent
between recording and replay. The condition is loaded from an {\tt is\_replay} global variable, so the
register holding the condition is different over a very short span of instructions (and explicity cleared afterwards).

Another tricky issue is handling system call memory parameters that are ``in-out''. During
recording we must copy the input buffer to the trace buffer, pass the system call a
pointer to the trace buffer, then copy the trace buffer contents back to the input
buffer. If we perform the first copy during replay, we'll overwrite the trace buffer values
holding the system call results we need. During replay we turn that copy into a no-op using
a conditional move to set the source address copy to the destination address.

We could allow the replay behavior of the interception library to diverge further from its
recording behavior, but it would have to be done very carefully. At least we'd have to ensure
the number of retired conditional branches was identical along both paths, and that register values
were consistent whenever we exit the interception library or trap to \system{} within the
interception library. It's simplest to minimize the divergence.

\subsection{Optimizing Reads With Block Cloning}

When an input file is on the same filesystem as the recorded trace and the filesystem supports copy-on-write cloning of file blocks, for large block-aligned {\tt read}s the system call buffering code clones the data to a per-thread ``cloned-data'' trace file, bypassing the normal system-call recording logic. This greatly reduces space and time overhead for file-read-intensive workloads; see the next section.

This optimization works by cloning the input blocks and then reading the input data from the original input file. This opens up a possible race: between the clone and the read, another process could overwrite the input file data, in which case the data read during replay would differ from the data read during recording, causing replay to fail. However, when a file read races with a write under Linux, the reader can receive an arbitrary mix of old and new data, so such behavior would almost certainly be a severe bug, and in practice such bugs do not seem to be common. The race could be avoided by reading from the cloned-data file instead of the original input file, but that performs very poorly because it defeats Linux's readahead optimizations (since the data in the cloned-data file is never available until just before it's needed).

\section{Results} \label{results}

Our results address the following questions:
\begin{itemize}
\item What is the run-time overhead of \system{} recording and replay, across different kinds of workloads?
\item How much of the overhead is due to the single-core limitation?
\item What are the impacts of the system-call buffering and block cloning optimizations?
\item What is the impact of not having to instrument code?
\item How much space do \system{} traces consume?
\end{itemize}

\subsection{Workloads}

We present a variety of workloads to illuminate \system{}'s strengths and weaknesses. Benchmarks were chosen to fit in system memory (to minimize the impact of I/O on test results), and to run for about 30 seconds each (except for \emph{cp} where a 30s run time would require it to not fit in memory).

\emph{cp} duplicates a {\tt git} checkout of {\tt glibc} (revision 2d02fd07) using {\tt cp -a} (15200 files constituting 732MB of data, according to {\tt du -h}). {\tt cp} is single-threaded, making intensive use of synchronous reads and a variety of other filesystem-related system calls.

\emph{make} builds DynamoRio \cite{Bruening2012} (version 6.1.0) with {\tt make -j8} ({\tt -j8} omitted when restricting to a single core). This tests potentially-parallel execution of many short-lived processes.

\emph{octane} runs the Google Octane benchmark under the Mozilla Spidermonkey Javascript engine (Mercurial revision 9bd900888753). This illustrates performance on CPU-intensive code in a complex language runtime.

\emph{htmltest} runs the Mozilla Firefox HTML forms tests (Mercurial revision 9bd900888753). The harness is excluded from recording (using {\tt mach mochitest -f plain --debugger \system{} dom/html/test/forms}). This is an example from real-world usage. About 30\% of user-space CPU time is in the harness.

\emph{sambatest} runs a Samba (git revision 9ee4678b) UDP echo test via {\tt make test TESTS=samba4.echo.udp}. This is an example from real-world usage.

All tests run on a Dell XPS15 laptop with a quad-core Intel Skylake CPU (8 SMT threads), 16GB RAM and a 512GB SSD using Btrfs in Fedora Core 23 Linux.

\subsection{Overhead}

\begin{table*}[t]
  \centering
  \begin{tabular}{lrrrrrrrr}
Workload & \shortstack[r]{Baseline \\ duration} & Record & Replay & \shortstack[r]{Single \\ core} & \shortstack[r]{Record \\ no-syscallbuf} & \shortstack[r]{Replay \\ no-syscallbuf} & \shortstack[r]{Record \\ no-cloning} & \shortstack[r]{DynamoRio- \\ null} \\
\hline
cp & 1.04s & 1.49$\times$ & 0.72$\times$ & 0.98$\times$ & 24.53$\times$ & 15.39$\times$ & 3.68$\times$ & 1.24$\times$ \\
make & 20.99s & 7.85$\times$ & 11.93$\times$ & 3.36$\times$ & 11.32$\times$ & 14.36$\times$ & 7.84$\times$ & 10.97$\times$ \\
octane & 32.73s & 1.79$\times$ & 1.56$\times$ & 1.36$\times$ & 2.65$\times$ & 2.43$\times$ & 1.80$\times$ & crash \\
htmltest & 23.74s & 1.49$\times$ & 1.01$\times$ & 1.07$\times$ & 4.66$\times$ & 3.43$\times$ & 1.50$\times$ & 14.03$\times$ \\
sambatest & 31.75s & 1.57$\times$ & 1.23$\times$ & 0.95$\times$ & 2.23$\times$ & 1.74$\times$ & 1.57$\times$ & 1.43$\times$ \\
  \end{tabular}
  \caption{Run-time overhead}
  \label{run-time-overhead-table}
\end{table*}

Table \ref{run-time-overhead-table} shows the wall-clock run time of various configurations, normalized to the run time of the baseline configuration. For \emph{octane}, because the benchmark is designed to run for a fixed length of time and report a score, we report the ratio of the baseline score to the configuration-under-test score instead --- except for replay tests, where the reported score will necessarily be the same as the score during recording. For \emph{octane} replay tests, we report the ratio of the baseline score to the recorded score, multiplied by the ratio of replay run time to recording run time.
Each test was run six times, discarding the first result and reporting the geometric mean of the other five results. Thus the results represent warm-cache performance.

``Single core'' reports the overhead of just restricting all threads to a single core using Linux {\tt taskset}.

``Record no-syscallbuf'' and ``Replay no-syscallbuf'' report overhead with the system-call buffering optimization disabled (which also disables block cloning). ``Record no-cloning'' reports overhead with just block cloning disabled.

``DynamoRio-null'' reports the overhead of running the tests under the DynamoRio \cite{Bruening2012} (version 6.1.0, the latest stable version at time of writing) ``null tool'', to estimate a lower bound for the overhead of using dynamic code instrumentation as an implementation technique. (DynamoRio is reported to be among the fastest dynamic code instrumentation engines.)

\subsection{Observations}

Overhead on \emph{make} is significantly higher than for the other workloads. Just forcing \emph{make} onto a single core imposes major slowdown. Also, \emph{make} creates many short-lived processes; it forks and execs 2430 processes. (The next most prolific workload is \emph{sambatest} with 89.) Our system-call buffering optimization only starts working in a process once \system{}'s preload library has been loaded, but typically at least 80 system calls are performed before that completes, so the effectiveness of system-call buffering is limited with short-lived processes.

\begin{figure}[t]
\includegraphics[scale=0.4]{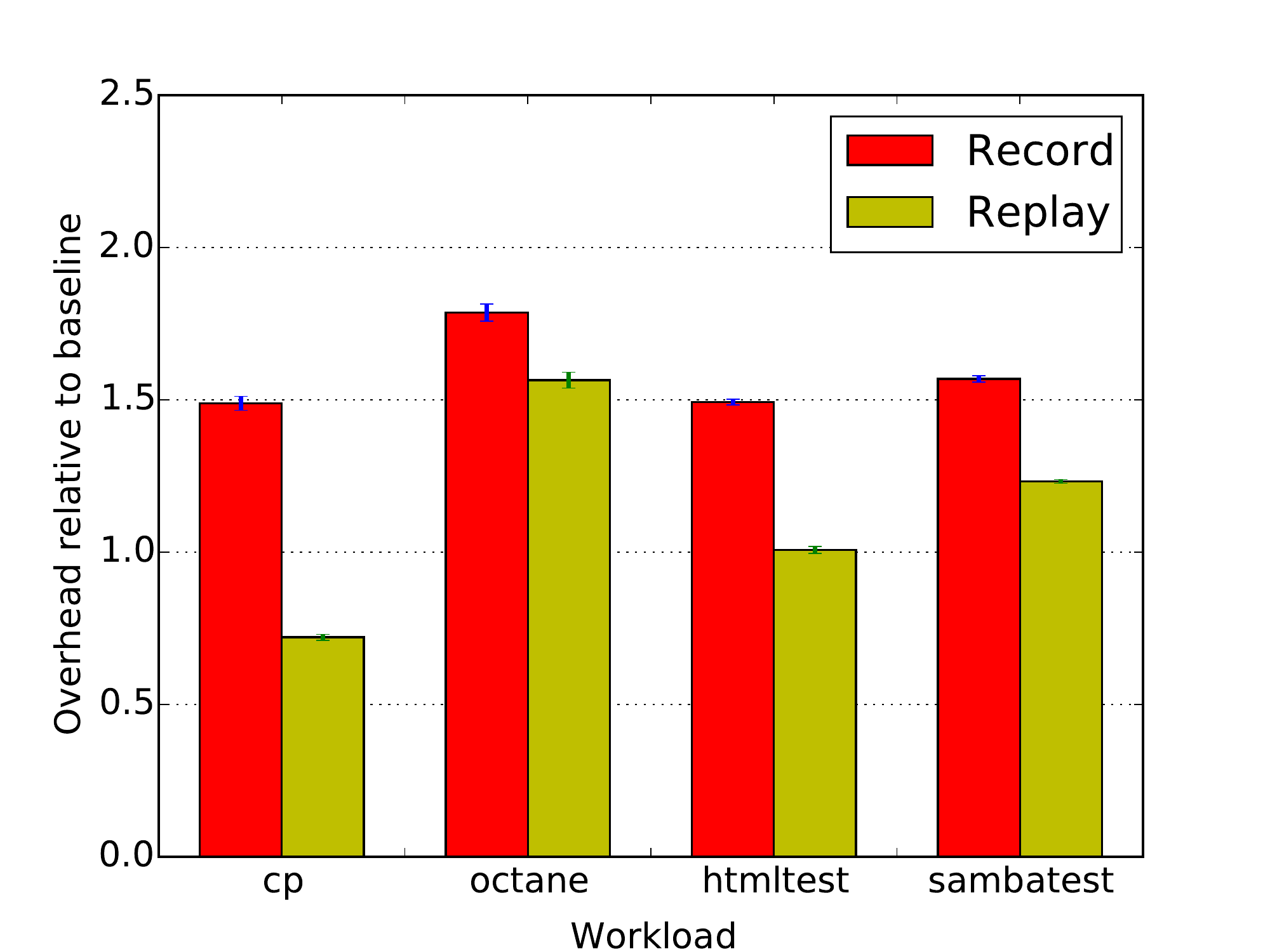}
\caption{Run-time overhead excluding \emph{make}}
\centering
\label{run-time-overhead-chart}
\end{figure}

Figure \ref{run-time-overhead-chart} shows the overall recording and replay overhead for workloads other than \emph{make}. Error bars in figures show 95\% confidence intervals; these results are highly stable across runs.

Excluding \emph{make}, \system{}'s recording slowdown is less than a factor of two. Excluding \emph{make}, \system{}'s replay overhead is lower than its recording overhead. Replay can even be faster than normal execution, in \emph{cp} because system calls do less work. For interactive applications, not represented here, replay can take much less time than the original execution because idle periods are eliminated.

\emph{octane} is the only workload here other than \emph{make} making significant use of multiple cores, and this accounts for the majority of \system{}'s overhead on \emph{octane}.

\begin{figure}[t]
\includegraphics[scale=0.4]{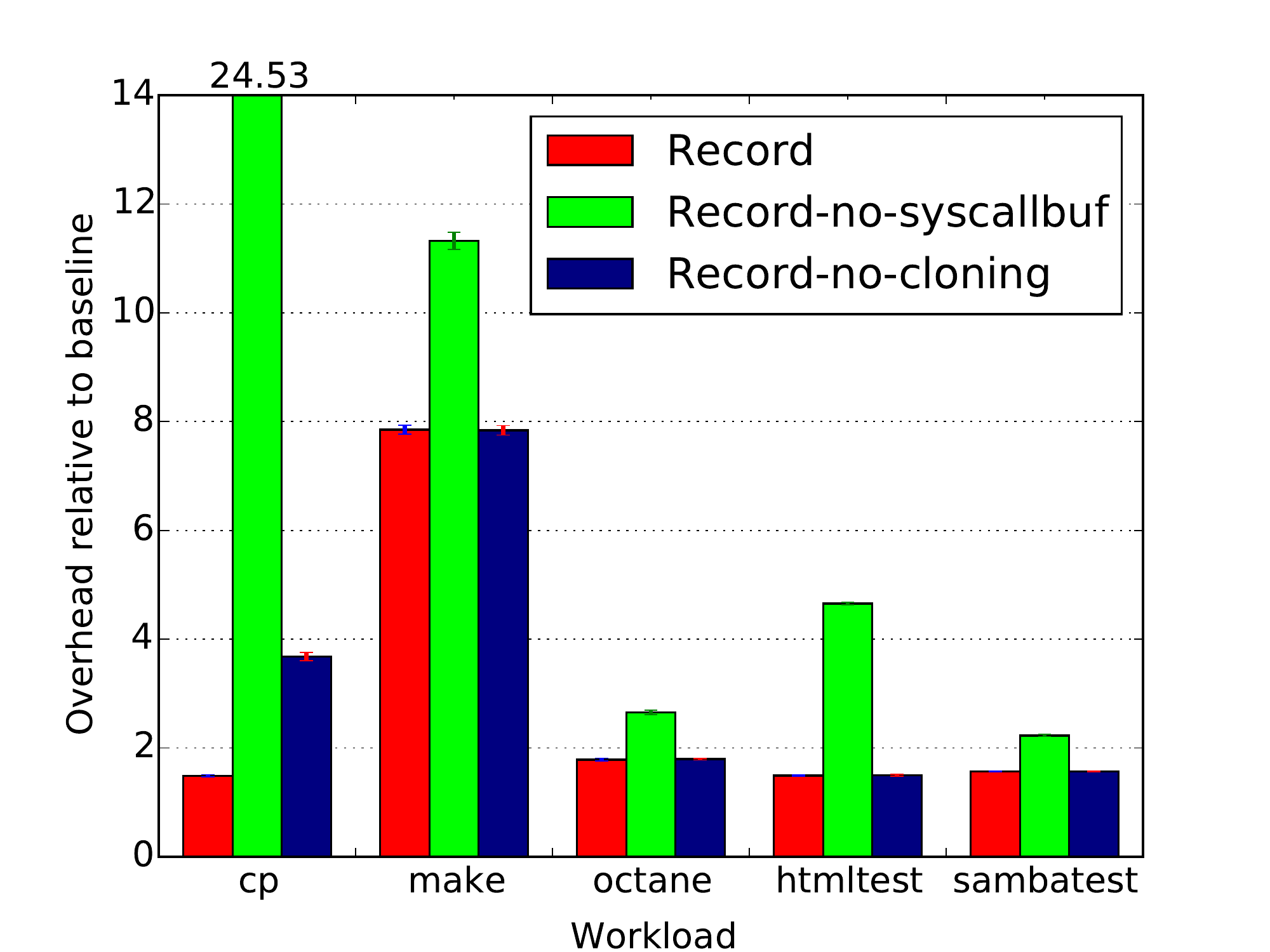}
\caption{Impact of optimizations}
\centering
\label{optimizations-chart}
\end{figure}

Figure \ref{optimizations-chart} shows the impact of system-call buffering and blocking cloning on recording. The system-call buffering optimization produces a large reduction in recording (and replay) overhead. Cloning file data blocks is a major improvement for \emph{cp} recording but has essentially no effect on the other workloads.

\begin{figure}[t]
\includegraphics[scale=0.4]{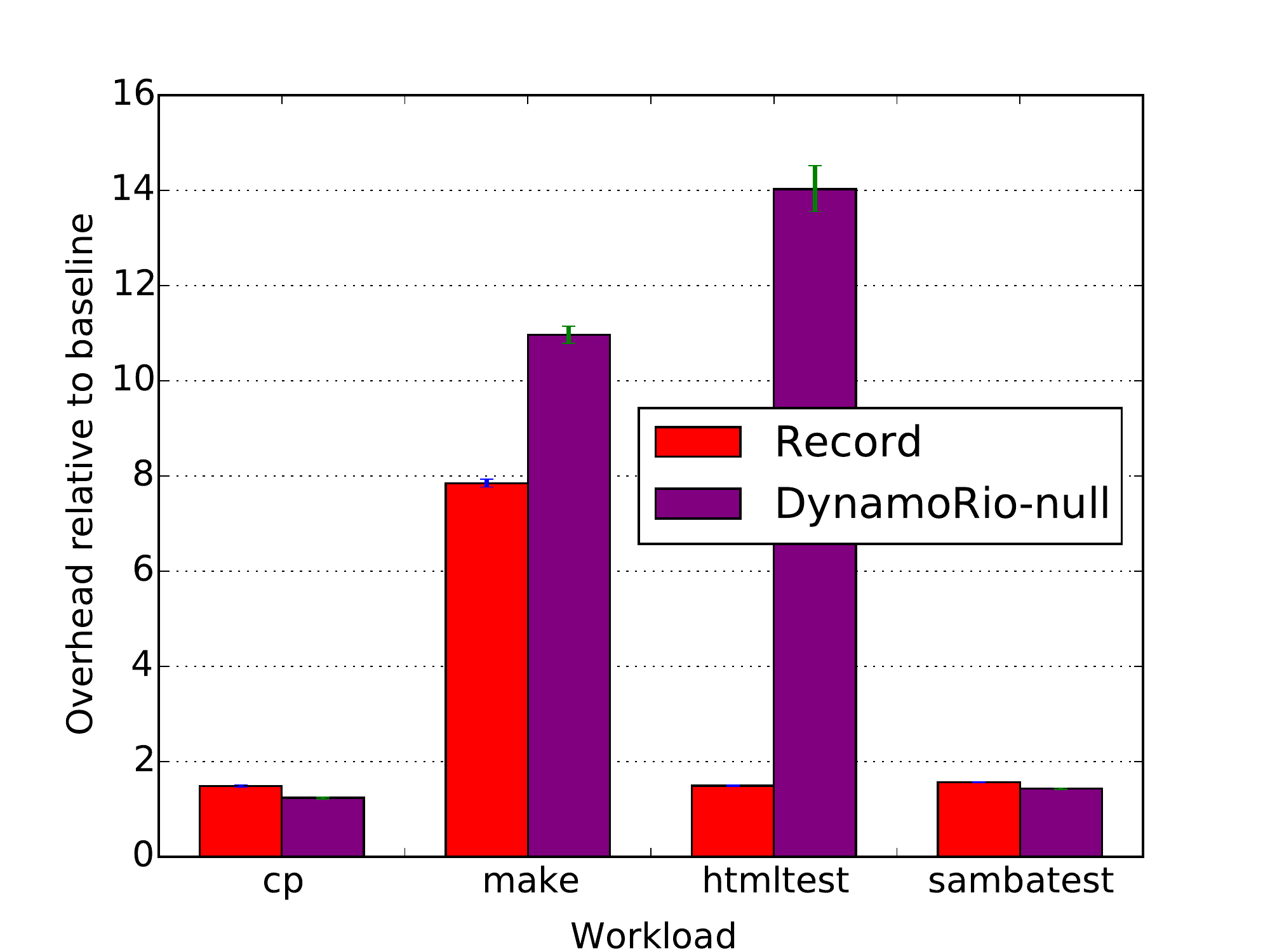}
\caption{Comparison with DynamoRio-null}
\centering
\label{dynamorio-chart}
\end{figure}

Figure \ref{dynamorio-chart} compares \system{} recording overhead with the overhead of DynamoRio's ``null tool'', which runs all code through the DynamoRio instrumentation engine but does not modify the code beyond whatever is necessary to maintain supervised execution; this represents a minimal-overhead code instrumentation configuration. DynamoRio crashed on \emph{octane}
\footnote{We reported DynamoRio's crash on our ``octane'' workload to the developers at \url{https://github.com/DynamoRIO/dynamorio/issues/1930}.} \emph{cp} executes very little user-space code and DynamoRio's overhead is low on that workload. On \emph{make} and \emph{sambatest} DynamoRio overhead is similar to \system{} recording, even though on \emph{make} DynamoRio can utilize multiple cores. On \emph{htmltest} DynamoRio's overhead is very high, possibly because that test runs a lot of Javascript with dynamically generated and modified machine code. Implementing record-and-replay on top of dynamic instrumentation would incur significant additional overhead, so we would expect the resulting system to have significantly higher overhead than \system{}.

\subsection{Storage Space Usage} \label{space-results}

\system{} traces contain three kinds of data:
\begin{itemize}
\item Cloned (or hard-linked) files used for memory-map operations
\item Cloned file blocks
\item All other trace data, especially event metadata and the results of general system calls
\end{itemize}

Memory-mapped files are almost entirely just the executables and libraries loaded by tracees. As long as the original files don't change and are not removed, which is usually true in practice, their clones take up no additional space and require no data writes. For simplicity, \system{} makes no attempt to eliminate duplicate file clones, so most traces contain many duplicates and reporting meaningful space usage for these files is both difficult and unimportant in practice.

\begin{table}[t]
  \centering
  \begin{tabular}{lrrr}
Workload & \shortstack[r]{Compressed \\ MB/s} & \shortstack[r]{{\tt deflate} \\ ratio} & \shortstack[r]{Cloned blocks \\ MB/s} \\
\hline
cp & 19.03 & 4.87$\times$ & 586.14 \\
make & 15.82 & 8.32$\times$ & 5.50 \\
octane & 0.08 & 8.33$\times$ & 0.00 \\
htmltest & 0.79 & 5.94$\times$ & 0.00 \\
sambatest & 6.85 & 21.87$\times$ & 0.00 \\
  \end{tabular}
  \caption{Storage space usage}
  \label{space-usage-table}
\end{table}

Table \ref{space-usage-table} shows the space usage of each workload, in megabytes per second, for the general trace data and the cloned file blocks. We compute the geometric mean of the data usage for each trace and divide by the run-time of the baseline configuration of the workload. The space consumed by each trace shows very little variation between runs.

Like cloned files, cloned file blocks do not consume space as long as the underlying data they're cloned from persists.

The different workloads have highly varying space consumption rates, but several megabytes per second is easy for modern systems to handle. Compression ratios vary depending on the kind of data being handled by the workload. In any case, in real-world usage trace storage has not been a concern.

\subsection{Memory Usage}

\begin{table}[t]
  \centering
  \begin{tabular}{lrrrr}
Workload & Baseline & Record & Replay & Single core \\
\hline
cp & 0.51 & 34.54 & 9.11 & 0.51 \\
make & 510.51 & 327.19 & 314.16 & 288.65 \\
octane & 419.47 & 610.48 & 588.01 & 392.95 \\
htmltest & 690.81 & 692.06 & 324.71 & 689.75 \\
sambatest & 298.68 & 400.49 & 428.79 & 303.03 \\
  \end{tabular}
  \caption{Memory usage (peak PSS MB)}
  \label{mem-usage-table}
\end{table}

\begin{figure}[t]
\centering
\includegraphics[scale=0.4]{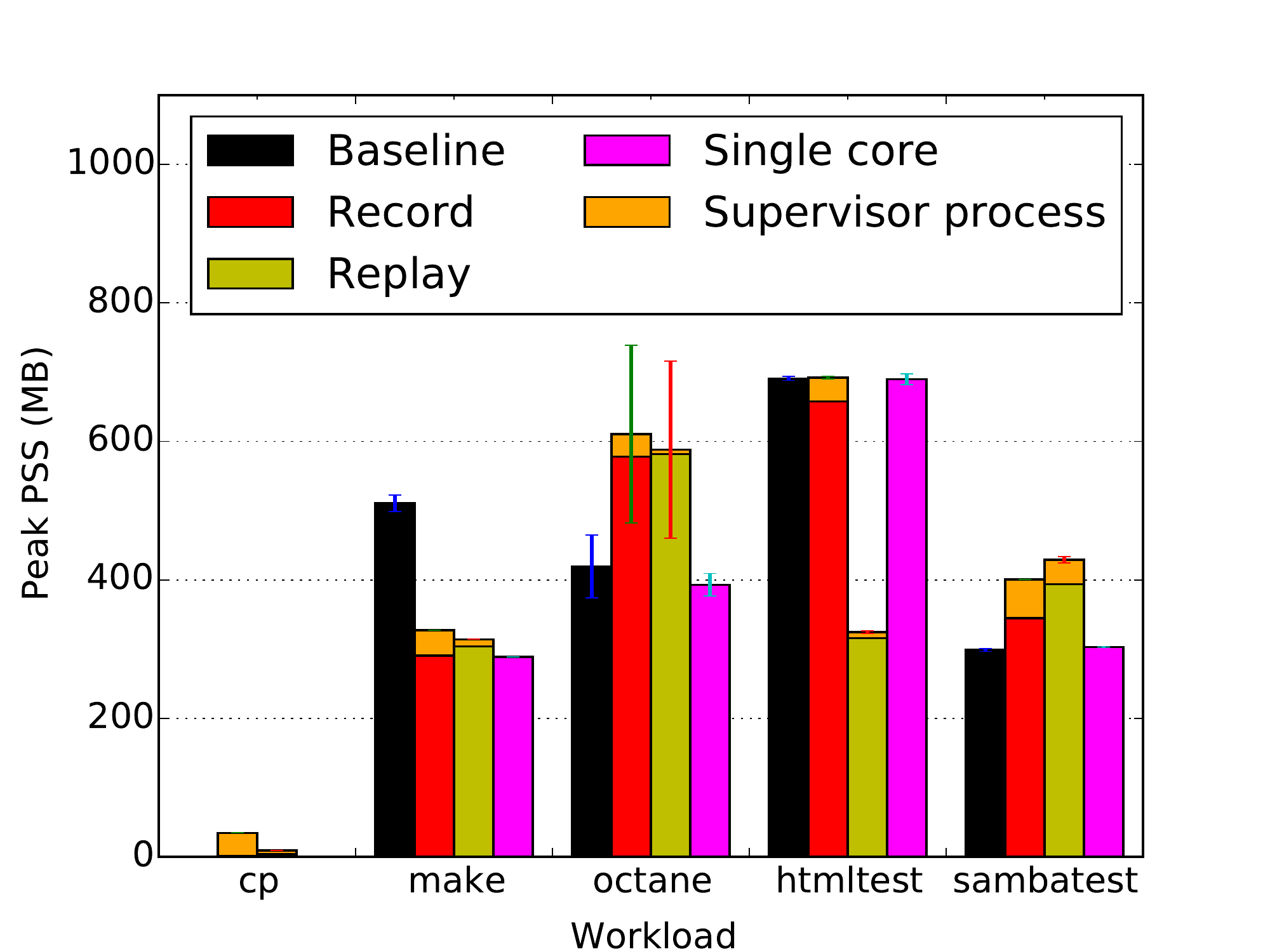}
\caption{Memory usage}
\label{mem-usage-chart}
\end{figure}

Table \ref{mem-usage-table} shows the memory usage of each workload. Every 10ms we sum the proportional-set-size (``PSS'') values of all workload processes (including \system{} if running); we determine the peak values for each run and take their geometric mean. In Linux, each page of memory mapped into a process's address space contributes $1/n$ pages to that process's PSS, where $n$ is the number of processes mapping the page; thus it is meaningful to sum PSS values over processes which share memory. The same data are shown in Figure \ref{mem-usage-chart}. In the figure, the fraction of PSS used by the \system{} process is shown in orange. Memory usage data was gathered in separate runs from the timing data shown above, to ensure the overhead of gathering memory statistics did not impact those results.

Given these experiments ran on an otherwise unloaded machine with 16GB RAM and all data fits in cache, none of these workloads experienced any memory pressure. \emph{cp} uses almost no memory. In \emph{make}, just running on a single core reduces peak PSS significantly because not as many processes run simultaneously. In \emph{octane} memory usage is volatile (highly sensitive to small changes in GC behavior) but recording significantly increases application memory usage; recording also increases application memory usage a small amount in \emph{sambatest} but slightly decreases it in \emph{htmltest}. (We would expect to see a small increase in application memory usage due to system-call buffers and scratch buffers mapped into application processes.) These effects are difficult to explain due to the complexity of the applications, but could be due to changes in timing and/or effects on application or operating system memory management heuristics.

Replay memory usage is similar to recording except in \emph{htmltest}, where it's dramatically lower because we're not replaying the test harness.

The important point is that \system{}'s memory overhead is relatively modest and is not an issue in practice.

\section{Hardware and Software Design Constraints} \label{constraints}

\subsection{Peformance Counters}

As discussed in Section \ref{counters}, \system{} uses hardware performance
counters to measure application progress. We need a counter that
increments very frequently so that the pair (counter, general purpose registers)
uniquely identifies a program execution point in practice, and it must be ``deterministic'' as described earlier.
Performance counters are typically used for statistical analysis, which does not
need this determinism property, so unsurprisingly it does not hold
for most performance counter implementations \cite{Weaver2013}. We are indeed fortunate that there is one satisfactory counter in Intel's modern Core models. The ideal performance counter for
our purposes would count the exact number of instructions retired as observed in user-space
(e.g., counting an instruction interrupted by a page fault just once); then the counter value
alone would always uniquely identify an execution point.

Being able to run \system{} in virtual machines is helpful for deployability.
KVM, VMware and Xen support virtualization of performance counters, and \system{}
is known to work in KVM and VMware (Xen not tested). Interestingly the VMware
\emph{VM exit clustering} optimization \cite{Agesen2012},
as implemented, breaks the determinism of the retired-conditional-branch
counter and must be manually disabled. Given a pair of close-together {\tt CPUID} instructions
separated by a conditional branch, the clustering optimization translates the code to
an instruction sequence that runs entirely in the host, reducing the number of VM exits from two (one per {\tt CPUID}) to one,
but performance counter effects are not emulated so the conditional branch is not counted.
In some runs of the instruction sequence the clustering optimization does not take effect
(perhaps because of a page fault?), so the count of retired conditional branches varies.

\subsection{Nondeterminisic Instructions}

Most CPU instructions are deterministic. As discussed in Section \ref{instructions},
the nondeterministic x86(-64) instructions can currently be handled or
(in practice) ignored, but this could change, so we need some improvements in this area.

Extending Linux (and hardware, where necessary) to support trapping on executions of
{\tt CPUID}, as we can for {\tt RDTSC}, would let us mask off capability bits so that
well-behaved software does not use the newer non-deterministic instructions.
Trapping and emulating {\tt CPUID} is also necessary to support trace portability
across machines with different CPU models and would let us avoid
pinning recording and replay to a specific core.

We would like to support record-and-replay of programs using hardware transactional
memory ({\tt XBEGIN}/{\tt XEND}). This could be done efficiently if the hardware
and OS could be configured to raise a signal on any failed transaction.

Trapping on all other nondetermnistic instructions (e.g.\ {\tt RDRAND}) would
be useful.

Porting \system{} to ARM failed because all ARM atomic memory operations
use the ``load-linked/store-conditional'' approach, which is inherently nondeterminstic.
The conditional store can fail because of non-user-space-observable activity, e.g.\ hardware interrupts, so counts of retired instructions or conditional branches for
code performing atomic memory operations are nondeterminstic. These
operations are inlined into very many code locations, so it appears
patching them is not feasible except via pervasive code instrumentation.
On x86(-64), atomic operations (e.g.\ compare-and-swap) are deterministic in terms of
user-space state, so there is no such problem.

\subsection{Use Of Shared Memory}

As noted in Section \ref{shmem}, \system{} does not support applications using
shared memory written to by non-recorded processes or kernel code --- unless there's some defined protocol for preventing read-after-write races that \system{} can observe, for example as is the case for Video4Linux.
Fortunately, popular desktop Linux frameworks (X11, PulseAudio) can be automatically configured per-client to minimize usage of shared memory to a few easily-handled edge cases.

\subsection{Operating System Features}

\system{} performance depends on leveraging relatively modern Linux features.
The {\tt seccomp-bpf} API, or something like it, is essential to selectively trapping system calls.
The Linux {\tt PERF\_COUNT\_SW\_CONTEXT\_SWITCHES} performance event is essential for handling blocking system calls.
The Linux copy-on-write file and block cloning APIs reduce overhead for I/O intensive applications.

\subsection{Operating System Design}

Any efficient record-and-replay system depends on clearly identifying a boundary within which code is replayed deterministically, and recording and replaying the timing and contents of all inputs into that boundary. In \system{}, that boundary is mostly the interface between the kernel and user-space. This works well with Linux: most of the Linux user/kernel interface is stable across OS versions, relatively simple, and well-documented, and it's easy to count hardware performance events occurring within the boundary (i.e.\ all user-space events for a specific process). This is less true in other operating systems. For example, in Windows, the user/kernel interface is not publicly documented, and is apparently more complex and less stable than in Linux. Implementing and maintaining the \system{} approach for an operating system like Windows would be considerably more challenging than for Linux, at least for anyone other than the OS vendor.

\section{Related Work}

\subsection{Whole-System Replay}

ReVirt \cite{Dunlap2002} was an early project that recorded and replayed the execution of an entire virtual machine. VMware \cite{Malyugin2007} used the same approach to support record-and-replay debugging in VMware Workstation, for a time, but discontinued the feature. The full-system simulator Simics has long supported reverse-execution debugging via deterministic reexecution \cite{Engblom2010}. There have been multiple independent efforts to add some amount of record-and-replay support to QEMU \cite{Dolan-Gavitt2015,Dovgalyuk2012,Srinivasan2011}. Whole-system record-and-replay can be useful, but for many users it is inconvenient to have to hoist the application into a virtual machine. Many applications of record-and-replay also require cheap checkpointing, and checkpointing an entire system is generally more expensive than checkpointing one or a few processes.

\subsection{Replaying User-Space With Kernel Support}

Scribe \cite{Laadan2010}, dOS \cite{Bergan2010} and Arnold \cite{Devecsery2014} replay a process or group of processes by extending the OS kernel with record-and-replay functionality. Requiring kernel changes makes maintenance and deployment more difficult --- unless record-and-replay is integrated into the base OS. But adding invasive new features to the kernel has risks, so if record-and-replay can be well implemented outside the kernel, moving it into the kernel may not be desirable.

\subsection{Pure User-Space Replay}

Approaches based on user-space interception of system calls have existed since at least MEC \cite{Chastain1999}, and later Jockey \cite{Saito2005} and liblog \cite{Geel2006}. Those systems did not handle asynchronous event timing. Also, techniques relying on library injection, such as liblog and Jockey, do not handle complex scenarios such as ASAN \cite{Serebryany2012} or Wine \cite{Wine} with their own preloading requirements. PinPlay \cite{Patil2010}, iDNA \cite{Bhansali2006}, UndoDB \cite{UndoDB} and TotalView ReplayEngine \cite{Gottbrath2008} use code instrumentation to record and replay asynchronous event timing. Unlike UndoDB and \system{}, PinPlay and iDNA instrument all loads, thus supporting parallel recording in the presence of data races and avoiding having to compute the effects of system calls, but this gives them higher overhead than the other systems.

As far as we know \system{} is the only pure user-space record-and-replay system to date that eschews code instrumentation in favour of hardware performance counters for asynchronous event timing, and it has the lowest overhead of the pure user-space systems.

\subsection{Higher-Level Replay}

Record-and-replay features have been integrated into language-level virtual machines. DejaVu \cite{Choi2001} added record-and-replay capabilities to the Jalape\~{n}o Java VM. Microsoft IntelliTrace \cite{Microsoft2013} instruments CLR bytecode to record high-level events and the parameters and results of function calls; it does not produce a full replay. Systems such as Chronon \cite{Deva2010} for Java instrument bytecode to collect enough data to provide the appearance of replaying execution for debugging purposes, without actually doing a replay. Dolos \cite{Burg2013} provides record-and-replay for JS applications in Webkit by recording and replaying the nondeterministic inputs to the browser. R2 \cite{Guo2008} provides record-and-replay by instrumenting library interfaces with assistance from the application developer. Such systems are all significantly narrower in scope than the ability to replay general user-space execution. 

\subsection{Parallel Replay}

Recording application threads running concurrently on multiple cores, with the possibility of data races, with low overhead, is extremely challenging. PinPlay \cite{Patil2010} and iDNA \cite{Bhansali2006} instrument shared-memory loads and report high overhead. SMP-ReVirt \cite{Dunlap2008} tracks page ownership using hardware page protection and reports high overhead on benchmarks with a lot of sharing. DoublePlay \cite{Veeraraghavan2011} runs two instances of the application and thus has high overhead when the application alone could saturate available cores. ODR \cite{Altekar2009} has low recording overhead but replay can be extremely expensive and is not guaranteed to reproduce the same program states.

The best hope for general low-overhead parallel recording seems to be hardware support. Projects such as FDR \cite{Xu2003}, BugNet \cite{Narayanasamy2005}, Rerun \cite{Hower2008}, DeLorean \cite{Montesinos2008} and QuickRec \cite{Pokam2013} have explored low-overhead parallel recording hardware.

\section{Future Work}

\system{} perturbs execution, especially by forcing all threads onto a single core, and therefore can fail to reproduce bugs that manifest outside \system{}. We have addressed this problem by introducing a ``chaos mode'' that intelligently adds randomness to scheduling decisions, enabling us to reproduce many more bugs, but that work is beyond the scope of this paper. There are many more opportunities to enhance the recorder to find more bugs.

Putting record-and-replay support in the kernel has performance benefits, e.g.\ reducing the cost of recording context switches. We may be able to find reusable primitives that can be added to kernels to improve the performance of user-space record-and-replay while being less invasive than a full kernel implementation.

Recording multiple processes running in parallel on multiple cores seems feasible if they do not share memory --- or, if they share memory, techniques inspired by SMP-ReVirt \cite{Dunlap2008} or dthreads \cite{Liu2011} may perform adequately for some workloads.

The applications of record-and-replay are perhaps more interesting and important than the base technology. For example, one can perform high-overhead dynamic analysis during replay \cite{Devecsery2014, Dolan-Gavitt2015, Patil2010}, potentially parallelized over multiple segments of the execution. With \system{}'s no-instrumentation approach, one could collect performance data such as sampled stacks and performance counter values during recording, and correlate that data with rich analysis generated during replay (e.g.\ cache simulation). Always-on record-and-replay would make finding and fixing bugs in the field much easier.

Supporting recording and replay of parallel shared-memory applications with potential races, while keeping overhead low, will be very difficult without hardware support. Demonstrating compelling applications for record-and-replay will build the case for including such support in commodity hardware.

\section{Conclusions}

The current state of Linux on commodity x86 CPUs enables single-core user-space record-and-replay with low overhead, without pervasive code instrumentation --- but only just. This is fortuitous; we use software and hardware features for purposes they were not designed to serve. It is also a recent development; five years ago {\tt seccomp-bpf} and the Linux file cloning APIs did not exist, and commodity architectures with a deterministic hardware performance counter usable from user-space had only just appeared (Intel Westmere)\footnote{Performance counters have been usable for kernel-implemented replay \cite{Dunlap2002,Olszewski2009} for longer, because kernel code can observe and compensate for events such as interrupts and page faults.}. By identifying the utility of these features for record-and-replay, we hope that they will be supported by an increasingly broad range of future systems. By providing an open-source, easy-to-deploy, production-ready record-and-replay framework we hope to enable more compelling applications of this technology.

\bibliography{Master}

\begin{thebibliography}{10}

\bibitem{UndoDB}
Reversible debugging tools for {C/C++} on {Linux} \& {Android}.
\newblock \url{http://undo-software.com}.
\newblock Accessed: 2016-04-16.

\bibitem{Microsoft2013}
Understanding {IntelliTrace} part {I}: What the @\#\$\% is {IntelliTrace}?
\newblock
  \url{https://blogs.msdn.microsoft.com/zainnab/2013/02/12/understanding-intellitrace-part-i-what-the-is-intellitrace}.
\newblock Accessed: 2016-04-16.

\bibitem{Wine}
Wine windows-on-posix framework.
\newblock \url{https://www.winehq.org}.
\newblock Accessed: 2016-09-20.

\bibitem{Agesen2012}
O.~Agesen, J.~Mattson, R.~Rugina, and J.~Sheldon.
\newblock Software techniques for avoiding hardware virtualization exits.
\newblock In {\em Proceedings of the 2012 USENIX Annual Technical Conference},
  June 2012.

\bibitem{Altekar2009}
G.~Altekar and I.~Stoica.
\newblock {ODR}: Output-deterministic replay for multicore debugging.
\newblock In {\em Proceedings of the ACM SIGOPS 22nd Symposium on Operating
  Systems Principles}, October 2009.

\bibitem{Bergan2010}
T.~Bergan, N.~Hunt, L.~Ceze, and S.~D. Gribble.
\newblock Deterministic process groups in {dOS}.
\newblock In {\em Proceedings of the 9th USENIX Symposium on Operating Systems
  Design and Implementation}, October 2010.

\bibitem{Bhansali2006}
S.~Bhansali, W.-K. Chen, S.~de~Jong, A.~Edwards, R.~Murray, M.~Drini\'{c},
  D.~Miho\v{c}ka, and J.~Chau.
\newblock Framework for instruction-level tracing and analysis of program
  executions.
\newblock In {\em Proceedings of the 2nd International Conference on Virtual
  Execution Environments}, June 2006.

\bibitem{Bruening2012}
D.~Bruening, Q.~Zhao, and S.~Amarasinghe.
\newblock Transparent dynamic instrumentation.
\newblock In {\em Proceedings of the 8th International Conference on Virtual
  Execution Environments}, March 2012.

\bibitem{Burg2013}
B.~Burg, R.~Bailey, A.~Ko, and M.~Ernst.
\newblock Interactive record/replay for web application debugging.
\newblock In {\em Proceedings of the 26th ACM Symposium on User Interface
  Software and Technology}, October 2013.

\bibitem{Chastain1999}
M.~E. Chastain.
\newblock \url{https://lwn.net/1999/0121/a/mec.html}, January 1999.
\newblock Accessed: 2016-04-16.

\bibitem{Choi2001}
J.-D. Choi, B.~Alpern, T.~Ngo, M.~Sridharan, and J.~Vlissides.
\newblock A perturbation-free replay platform for cross-optimized multithreaded
  applications.
\newblock In {\em Proceedings of the 15th International Parallel and
  Distributed Processing Symposium}, April 2001.

\bibitem{Deva2010}
P.~Deva.
\newblock
  \url{http://chrononsystems.com/blog/design-and-architecture-of-the-chronon-record-0},
  December 2010.
\newblock Accessed: 2016-04-16.

\bibitem{Devecsery2014}
D.~Devecsery, M.~Chow, X.~Dou, J.~Flinn, and P.~M. Chen.
\newblock Eidetic systems.
\newblock In {\em Proceedings of the 11th USENIX Symposium on Operating Systems
  Design and Implementation}, October 2014.

\bibitem{Dolan-Gavitt2015}
B.~Dolan-Gavitt, J.~Hodosh, P.~Hulin, T.~Leek, and R.~Whelan.
\newblock Repeatable reverse engineering with {PANDA}.
\newblock In {\em Proceedings of the 5th Program Protection and Reverse
  Engineering Workshop}, December 2015.

\bibitem{Dovgalyuk2012}
P.~Dovgalyuk.
\newblock Deterministic replay of system’s execution with multi-target {QEMU}
  simulator for dynamic analysis and reverse debugging.
\newblock 2012.

\bibitem{Dunlap2002}
G.~Dunlap, S.~King, S.~Cinar, M.~A. Basrai, and P.~M. Chen.
\newblock {ReVirt}: Enabling intrusion analysis through virtual-machine logging
  and replay.
\newblock In {\em Proceedings of the 5th USENIX Symposium on Operating Systems
  Design and Implementation}, December 2002.

\bibitem{Dunlap2008}
G.~W. Dunlap, D.~G. Lucchetti, M.~A. Fetterman, and P.~M. Chen.
\newblock Execution replay of multiprocessor virtual machines.
\newblock In {\em Proceedings of the 4th ACM SIGPLAN/SIGOPS International
  Conference on Virtual Execution Environments}, March 2008.

\bibitem{Engblom2012}
J.~Engblom.
\newblock A review of reverse debugging.
\newblock In {\em System, Software, SoC and Silicon Debug Conference},
  September 2012.

\bibitem{Engblom2010}
J.~Engblom, D.~Aarno, and B.~Werner.
\newblock Full-system simulation from embedded to high-performance systems.
\newblock In {\em Processor and System-on-Chip Simulation}, 2010.

\bibitem{Geel2006}
D.~Geels, G.~Altekar, S.~Shenker, and I.~Stoica.
\newblock Replay debugging for distributed applications.
\newblock In {\em Proceedings of the 2006 USENIX Annual Technical Conference},
  June 2006.

\bibitem{Gottbrath2008}
C.~Gottbrath.
\newblock Reverse debugging with the {TotalView} debugger.
\newblock In {\em Cray User Group Conference}, May 2008.

\bibitem{Guo2008}
Z.~Guo, X.~Wang, J.~Tang, X.~Liu, Z.~Xu, M.~Wu, M.~F. Kaashoek, and Z.~Zhang.
\newblock {R2}: An application-level kernel for record and replay.
\newblock In {\em Proceedings of the 8th USENIX Symposium on Operating Systems
  Design and Implementation}, December 2008.

\bibitem{Hower2008}
D.~Hower and M.~Hill.
\newblock {Rerun}: Exploiting episodes for lightweight memory race recording.
\newblock In {\em Proceedings of the 35th Annual International Symposium on
  Computer Architecture}, June 2008.

\bibitem{Holzle1991}
U.~Hölzle, C.~Chambers, and D.~Ungar.
\newblock Optimizing dynamically-typed object-oriented languages with
  polymorphic inline caches.
\newblock In {\em Proceedings of the 1991 European Conference on
  Object-Oriented Programming}, July 1991.

\bibitem{Laadan2010}
O.~Laadan, N.~Viennot, and J.~Nieh.
\newblock Transparent, lightweight application execution replay on commodity
  multiprocessor operating systems.
\newblock In {\em Proceedings of the ACM International Conference on
  Measurement and Modeling of Computer Systems}, June 2010.

\bibitem{Liu2011}
T.~Liu, C.~Curtsinger, and E.~Berger.
\newblock {Dthreads}: Efficient deterministic multithreading.
\newblock In {\em Proceedings of the ACM SIGOPS 23rd Symposium on Operating
  Systems Principles}, October 2011.

\bibitem{Malyugin2007}
V.~Malyugin, J.~Sheldon, G.~Venkitachalam, B.~Weissman, and M.~Xu.
\newblock {ReTrace}: Collecting execution trace with virtual machine
  deterministic replay.
\newblock In {\em Proceedings of the Workshop on Modeling, Benchmarking and
  Simulation}, June 2007.

\bibitem{Montesinos2008}
P.~Montesinos, L.~Ceze, and J.~Torrellas.
\newblock {DeLorean}: Recording and deterministically replaying shared-memory
  multiprocessor execution efficiently.
\newblock In {\em Proceedings of the 35th Annual International Symposium on
  Computer Architecture}, June 2008.

\bibitem{Narayanasamy2005}
S.~Narayanasamy, G.~Pokam, and B.~Calder.
\newblock Bugnet: Continuously recording program execution for deterministic
  replay debugging.
\newblock In {\em Proceedings of the 32nd Annual International Symposium on
  Computer Architecture}, June 2005.

\bibitem{Olszewski2009}
M.~Olszewski, J.~Ansel, and S.~Amarasinghe.
\newblock {Kendo}: Efficient deterministic multithreading in software.
\newblock In {\em Proceedings of the 14th International Conference on
  Architectural Support for Programming Languages and Operating Systems}, March
  2009.

\bibitem{Patil2010}
H.~Patil, C.~Pereira, M.~Stallcup, G.~Lueck, and J.~Cownie.
\newblock {PinPlay}: A framework for deterministic replay and reproducible
  analysis of parallel programs.
\newblock In {\em Proceedings of the 8th Annual IEEE/ACM International
  Symposium on Code Generation and Optimization}, April 2010.

\bibitem{Pokam2013}
G.~Pokam, K.~Danne, C.~Pereira, R.~Kassa, T.~Kranich, S.~Hu, J.~Gottschlich,
  N.~Honarmand, N.~Dautenhahn, S.~King, and J.~Torrellas.
\newblock {QuickRec}: Prototyping an intel architecture extension for record
  and replay of multithreaded programs.
\newblock In {\em Proceedings of the 40th Annual International Symposium on
  Computer Architecture}, June 2013.

\bibitem{Saito2005}
Y.~Saito.
\newblock Jockey: A user-space library for record-replay debugging.
\newblock In {\em Proceedings of the 6th International Symposium on Automated
  Analysis-driven Debugging}, September 2005.

\bibitem{Serebryany2012}
K.~Serebryany, D.~Bruening, A.~Potapenko, and D.~Vyukov.
\newblock Addresssanitizer: A fast address sanity checker.
\newblock In {\em Proceedings of the 2012 USENIX Annual Technical Conference},
  June 2012.

\bibitem{Srinivasan2011}
D.~Srinivasan and X.~Jiang.
\newblock Time-traveling forensic analysis of {VM}-based high-interaction
  honeypots.
\newblock In {\em Security and Privacy in Communication Networks: 7th
  International ICST Conference}, September 2011.

\bibitem{Veeraraghavan2011}
K.~Veeraraghavan, D.~Lee, B.~Wester, J.~Ouyang, P.~M. Chen, J.~Flinn, and
  S.~Narayanasamy.
\newblock Doubleplay: Parallelizing sequential logging and replay.
\newblock In {\em Proceedings of the 16th International Conference on
  Architectural Support for Programming Languages and Operating Systems}, March
  2011.

\bibitem{Weaver2013}
V.~Weaver, D.~Terpstra, and S.~Moore.
\newblock Non-determinism and overcount on modern hardware performance counter
  implementations.
\newblock In {\em Proceedings of the IEEE International Symposium on
  Performance Analysis of Systems and Software}, April 2013.

\bibitem{Xu2003}
M.~Xu, R.~Bodik, and M.~D. Hill.
\newblock A ``flight data recorder'' for enabling full-system multiprocessor
  deterministic replay.
\newblock In {\em Proceedings of the 30th Annual International Symposium on
  Computer Architecture}, June 2003.

\end{thebibliography}
\bibliographystyle{abbrv} 

\end{document}